\documentclass[%
 superscriptaddress,
 twocolumn,
nofootinbib,
 amsmath,amssymb,
 aps,
 prc,
]{revtex4-2}

\usepackage[colorlinks=true,allcolors=blue]{hyperref} 
\usepackage{graphicx}
\usepackage{dcolumn}
\usepackage{bm}
\usepackage{physics} 
\newcolumntype{d}[1]{D{.}{.}{#1}} 
\usepackage{xcolor}
\usepackage{soul}
\usepackage{amsthm}

\usepackage{ulem}
\usepackage{orcidlink}

\begin{document}


\title{Electron capture of superheavy nuclei with realistic lepton wave functions}

\author{A. Ravli\'c\,\orcidlink{0000-0001-9639-5382}}
\email[]{ravlic@frib.msu.edu}
\affiliation{Facility for Rare Isotope Beams, Michigan State University, East Lansing, Michigan 48824, USA}

\author{P. Schwerdtfeger\,\orcidlink{0000-0003-4845-686X}}
\affiliation{Centre for Theoretical Chemistry and Physics, The New Zealand Institute for Advanced Study (NZIAS), Massey University Albany, Private Bag 102904, Auckland 0745, New Zealand}

\author{W. Nazarewicz\,\orcidlink{0000-0002-8084-7425}}
\email[]{witek@frib.msu.edu}
\affiliation{Facility for Rare Isotope Beams, Michigan State University, East Lansing, Michigan 48824, USA}
\affiliation{Department of Physics and Astronomy, Michigan State University, East Lansing, Michigan 48824, USA}

\begin{abstract}
The superheavy nuclei push the periodic table of the elements and the chart of the nuclides to their limits, providing a unique laboratory for studies of the electron-nucleus interactions. The most important weak decay mode in known superheavy nuclei is electron capture (EC). In the standard calculations of EC, the lepton wave functions are usually considered in the lowest-order approximation. In this work, we investigate the sensitivity of EC rates on the choice of the electron wave functions by (i) assuming the single-particle approximation for the electron wave functions, and (ii)  carrying out Dirac-Hartree-Fock (DHF) calculations. The nuclear response is generated based on the state-of-the-art  quasiparticle random 
phase approximation employing relativistic nuclear energy density functional theory. We show that using the improved lepton wave functions reduces the EC rates up to 40\% in the superheavy nucleus oganesson ($Z=118$). Interestingly, because of screening effects,  the difference between the EC rates obtained with the DHF and single-particle calculations is fairly small.
\end{abstract}

\date{\today}

\maketitle

\section{Introduction}
 Electron capture (EC) is a process mediated by the weak nuclear interaction in which a nucleus absorbs a bound-state electron  and
  causes the emission of an electron neutrino \cite{Bambynek1977a}. It occurs in the proton-rich regions of the nuclear landscape and acts in the same direction as the $\beta^+$-decay. Both are important for astrophysical processes such as the $p$-process and the $rp$-process, responsible for the creation of proton-rich nuclei \cite{Schatz1998a,Arnould2003a}. In stellar environments, EC is the main mechanism responsible for driving the deleptonization and determining the dynamics of core-collapse supernovae \cite{Langanke2003a,Langanke2021a}. Since the $Q$-value of the EC is around 1\,MeV higher than that of  $\beta^+$-decay, EC generally dominates for nuclei closer to the valley of $\beta$-stability where the $\beta^+$-decay is suppressed. Moreover, because the phase-space of EC decay scales with the nuclear charge as $Z^3$, it becomes the prominent weak decay in proton-rich superheavy nuclei. Usually, the leading  decay modes of SHN are alpha-decay and fission \cite{Giuliani2018a,Giuliani2019a,Heenen2015a,Olsen2019a,Staszczak2013a}.  There are cases, however,  where 
  EC decays can compete with these modes \cite{Ravlic2024b,Smits2023a,Smits2024a,Khuyagbaatar2020}. Therefore, a proper theoretical description of this process is necessary as it can provide guidance for future experimental efforts in the SHN region. This requires a careful treatment of both the nuclear and  atomic aspect of the problem. However, since the binding energies of innermost bound electrons in superheavy atoms become comparable to the electron mass, relativistic effects come into play, which leads to many fascinating phenomena \cite{Jerabek2018a,Indelicato2007a,Smits2020a}. In addition, to precisely describe the chemistry of superheavy atoms one has to consider quantum electrodynamics (QED) effects, crucial in determining the ground-state configurations, together with electron-electron correlations \cite{Schwerdtfeger2015a,Malyshev2022a,Savelyev2023a,Mohr1998a}.

In the EC process, the electrons are captured from the bound orbitals, while the outgoing neutrinos are in the continuum. By neglecting a small neutrino mass, the neutrino radial wave functions are well described by the spherical Bessel functions \cite{Koshigiri1979a,Bambynek1977a,Horiuchi2021a}. On the other hand, to obtain the electron radial wave functions (ERWFs), one has to solve the bound-state Dirac equation with the actual nuclear electrostatic potential. The most ubiquitous approach is to expand the ERWFs in the powers of electron mass $m_e$, nuclear charge $Ze$, and nuclear radius $R$, and retain the lowest-order (LO) contribution \cite{Behrens1970a,Behrens1971a,Bambynek1977a}. This makes it possible to factor different terms with specific nuclear operators and  parametrize the radial dependence of the ERWF with its value close to the origin using the so-called \textit{Coulomb amplitudes} $\beta_x$. 

Calculations of EC rates in SHN performed up to date were made using rather simplified phenomenological mean-field  models \cite{Sarriguren2019a,Sarriguren2021a,Sarriguren2022a} or the microscopic-macroscopic models \cite{Moller1997a,Moller1994a}. The nuclear excitations were determined using the quasiparticle random-phase approximation (QRPA). However, these calculations considered contribution of only the allowed Gamow-Teller (GT) transitions. Furthermore, the ERWFs were limited to the LO approximation. By using a more sophisticated framework, based on the relativistic  density functional  theory (DFT) and QRPA, $\beta^+$/EC rates of specific proton-rich nuclei up to $Z = 50$ were calculated in Ref. \cite{Niu2013a}, also at the LO. Electron capture rates calculated going beyond the LO approximation were presented in Refs. \cite{Sevestrean2023a,Mougeot2019a}, but using a schematic nuclear model.  Only recently, in Ref. \cite{Horiuchi2021a},  the non-relativistic DFT+QRPA theory applied to $\beta^-$-decays of  nuclei up to tin ($Z = 50$) used ERWFs beyond the LO. It was found that the LO approximation overestimates the rates by 5--15\%.  We also note that the large-scale calculations of $\beta^-$-decay rates, used in astrophysical nucleosynthesis simulations, do employ the LO approximation to calculate the half-lives \cite{Ney2020a,Marketin2016a}. 
Let us also mention that in Ref. \cite{Pachucki2007} the radiative EC process has been studied. By relating it to the non-radiative EC rate, they avoided nuclear structure calculations.

In this work,  we employ the  ERWFs obtained by solving the  Dirac equation. As a first method, we assume that a single electron is bound in an orbital $x$.
To obtain the wave functions and the bound-state energies, we
use the single-particle Dirac solver RADIAL \cite{Salvat1995a}. However, this is a crude approximation as it neglects electron-electron interaction and electron screening effects. To this end, we also obtain the ERWFs using many-body Dirac-Hartree-Fock (DHF) calculations within the numerical GRASP program package \cite{GRASP}. The nuclear ground state is obtained within relativistic DFT theory applying the point-coupling energy density functional (EDF)  DD-PC1 \cite{Niksic2008a}. We assume axially-deformed and reflection-symmetric nuclear shapes. The excited states and the nuclear response are generated with the relativistic QRPA in the proton-neutron channel (pnRQRPA), developed in Refs. \cite{Ravlic2021a,Ravlic2024a}. This framework allows for efficient quantified global predictions of weak interaction rates  \cite{Ravlic2024b}. 

This paper is structured  as follows. In Sec. \ref{sec:theory}, we present the EC rate expressions originating from the current-current form of the weak-interaction Hamiltonian. A method on how to calculate required matrix elements within the nuclear linear response pnRQRPA is developed in terms of transition densities. Subsequently, the LO reduction is provided in Appendix \ref{sec:appa}. In Sec. \ref{sec:results} we present calculations of EC rates in the medium-mass Fe, Pd, and Dy isotopic chains, as well as in the superheavy Og chain, comparing the LO limit of ERWFs with more advanced single-particle Dirac (DWF) and DHF (DHFWF) wave functions.

\section{Theoretical formalism}\label{sec:theory}
Here we present detailed expressions for the EC rate using ERWFs obtained from advanced atomic solvers. The derivation follows Refs. \cite{Horiuchi2021a,Koshigiri1979a}, originally developed for $\beta^\pm$-decay, and extends expressions to the particular case of EC by considering bound-state electron wave functions. To calculate the required matrix elements, we extend the formalism of the linear response pnRQRPA to include improved lepton wave functions. For completeness, we perform the reduction to the LO limit and obtain results consistent with the existing literature \cite{Bambynek1977a,Behrens1970a,Behrens1971a}.

\subsection{Electron capture using exact ERWFs}
To describe the EC process, the electron is assumed to be in a bound orbital $x$ with energy $E_x$ and the two-component  Dirac wave function
\begin{equation}\label{eq:electron_wavefunction}
    \Psi_{e^-}(\boldsymbol{r}) = \begin{pmatrix}
        G_{\kappa_x}(\boldsymbol{r}) \chi_{\kappa_x \mu_x} \\
        i F_{\kappa_x}(\boldsymbol{r}) \chi_{-\kappa_x \mu_x}
    \end{pmatrix},
\end{equation}
where 
\begin{equation}
\chi_{\kappa_x \mu_x} = [Y_{l_x} \otimes \xi_{1/2}]_{j_x \mu_x} = \sum \limits_{m_x s_x} C^{j_x \mu_x}_{l_x m_x 1/2 s_x} Y_{l_x m_x} \xi_{1/2 s_x},
\end{equation}
is the angular part of the wave function, with $j_x$ the total angular momentum, $l_x$ the orbital angular momentum, and $\xi_{1/2 s_x}$ the spin-1/2 wave function, coupled by the Clebsch-Gordan coefficient $C^{j_x \mu_x}_{l_x m_x 1/2 s_x}$.  As usual, the relativistic block number $\kappa_x$ is defined as
\begin{equation}
    \kappa_x = \left\{ \begin{array}{c}
         l_x, \quad j_x = l_x - 1/2  \\
         -(l_x + 1), \quad j_x = l_x + 1/2.
    \end{array} \right.
\end{equation}
Boldspace notation is used to denote vectors in the coordinate space.

Assuming a single-particle problem, the electron radial wave functions $G_{\kappa_x}, F_{\kappa_x}$ for orbital $x \equiv (l_x, j_x)$ are obtained by solving the Dirac equation
\begin{align}
    \begin{split}
        \frac{d G_{\kappa_x}}{dr} &= \frac{\kappa_x}{r}G_{\kappa_x} - [E_x - V_{\rm nuc}(r) + 2m_e] F_{\kappa_x}, \\
        \frac{d F_{\kappa_x}}{dr} & = [E_x - V_{\rm nuc}(r)]G_{\kappa_x}  + \frac{\kappa_x}{r} F_{\kappa_x}, \\
    \end{split}
\end{align}
where $V_{\rm nuc}(r)$ is the Coulomb potential generated by a spherical nuclear charge distribution. 

By taking into account all electrons in the atomic cloud, we can also study the influence of electron-electron interactions on the bound-state wave function. A configuration state function (CSF), $|\gamma \Pi J M \rangle$, is a coupled linear combination of $Z$ Slater determinants, that has a well-defined parity $\Pi$, total angular momentum $J$ and its projection $M$, with $\gamma$ denoting other quantum numbers. The total Hamiltonian of the system of $Z$ electrons is written as:
\begin{equation}\label{eq:electron_hamiltonian}
    \hat{H} = \hat{H}_{\rm D} + \hat{V}_{\rm nuc} +  \sum \limits_{i = 1}^{Z-1} \sum \limits_{j = i+1}^Z \frac{1}{|\boldsymbol{r}_i - \boldsymbol{r}_j|},
\end{equation}
where the Dirac Hamiltonian is
\begin{equation}
    \hat{H}_{\rm D} = \sum \limits_{i = 1}^Z \gamma_0\boldsymbol{\gamma} \cdot \hat{\boldsymbol{\partial}}_i + \gamma_0 m_e,
\end{equation}
$\hat{V}_{\rm nuc}$ is the nuclear charge potential, and the last term represents the Coulomb interaction between the electrons. Further corrections originate from Breit interaction and QED effects \cite{Schwerdtfeger2015a}. To find the eigenstates of (\ref{eq:electron_hamiltonian}), we employ the numerical atomic structure package GRASP \cite{GRASP}. The GRASP calculations presented in this work utilize only one CSF and are therefore solved by the variational DHF method.

The outgoing neutrino moving with the linear momentum   $\boldsymbol{k}_\nu$ is embedded in the continuum, and its wave function can be expanded in spherical partial waves \cite{Horiuchi2021a}:
\begin{align}\label{eq:neutrino_multipoles}
\begin{split}
    u_{s_\nu}(\boldsymbol{k}_\nu) e^{i \boldsymbol{k}_\nu \cdot \boldsymbol{r}} &= \sum \limits_{\kappa_\nu m_\nu \mu_\nu} \frac{4\pi}{\sqrt{2}} Y^*_{l_{\kappa_\nu} m_\nu} C^{j_{\kappa_\nu} \mu_\nu}_{l_{\kappa_\nu} m_\nu 1/2 s_\nu} \\
    &\times \begin{pmatrix}
        g_{\kappa_\nu} \chi_{\kappa_\nu \mu_\nu} \\
        i f_{\kappa_\nu} \chi_{-\kappa_\nu \mu_\nu}
    \end{pmatrix},
\end{split}
\end{align}
where $\kappa_\nu$ is the neutrino relativistic block number, and $\chi_{\kappa_\nu \mu_\nu} = [Y_{l_\nu} \otimes \xi_{1/2}]_{j_\nu \mu_\nu}$, with total(orbital) angular momentum $j_\nu$($l_\nu$) and spin-1/2 wave function $\xi_{1/2 s_\nu}$.  The neutrino radial wave functions are  the solutions of the free spherical Dirac equation:
\begin{equation}
    g_\kappa(r) = j_{l_\kappa}(k_\nu r), \quad f_\kappa(r) = S_\kappa j_{\bar{l}_\kappa}(k_\nu r),
\end{equation}
with $\bar{l}_\kappa = l_{-\kappa}$, $S_\kappa = \text{sgn}(\kappa)$, and $j_{l_\kappa}$ being the spherical Bessel function.

To obtain the EC rate, we define the weak-interaction Hamiltonian in the Fermi current-current approximation and apply Fermi's golden rule \cite{Koshigiri1979a}. The expansion of the neutrino continuum wave function, and coupling the angular momentum, greatly simplify the final expression, as detailed in Refs. \cite{Koshigiri1979a,Horiuchi2021a} for $\beta^\pm$-decays. In this work, however, the kinematics has to be adjusted to the special case of EC. The  final expression for the EC rate from an electron orbital $x$ is:
\begin{align}\label{eq:rate_equation}
\begin{split}
    \lambda_x &= \frac{G_F^2 V_{ud}^2}{2\pi} \sum \limits_{i,f} (E_0^{(i,f)} + E_x)^2 \\
    &\times \sum \limits_{\kappa_\nu} \sum \limits_{JL} \frac{1}{2J_i + 1} \left| \langle f || \Xi_{JL}(\kappa_x, \kappa_\nu) || i \rangle \right|^2,
\end{split}
\end{align}
where $E_0^{(i,f)}$ is the end-point energy, {i.e.}, the difference between the energy of initial and final combined nuclear and lepton state $E_0^{(i,f)} = E_i - E_f$, where $J_i(J_f)$ is the angular momentum of initial(final) state. $G_F$ is the Fermi constant and $V_{ud}$ is the up-down mixing angle of the CKW matrix. The matrix element $\langle f || \Xi_{JL}(\kappa_x, \kappa_\nu) || i \rangle$ contains the nuclear-atomic coupling. By introducing the definition of a transition density $\delta \rho^{(i,f)}_{F_{JL}}$ induced by the external field $\hat{F}_{JL}$ as
\begin{equation}\label{eq:transition_density}
   \langle i || \hat{F}_{JL} || f \rangle  = \int d \boldsymbol{r} \delta \rho^{(i,f)}_{F_{JL}} (\boldsymbol{r}) , 
\end{equation}
for a transition between initial state $i$ and final state $f$, the EC matrix element becomes:
\begin{widetext}
\begin{align}\label{eq:matrix_element_ERWF}
    \begin{split}
        \langle i || \Xi_{J L}(\kappa_x, \kappa_\nu) || f \rangle &= \int d\boldsymbol{r} \left\{ -g_V \sqrt{\frac{2J+1}{4 \pi}} \delta \rho^{(i,f)}_{\boldsymbol{C}_J}(\boldsymbol{r}) [g_{\kappa_\nu} G_{\kappa_x} S_{0 JJ}(\kappa_\nu, \kappa_x) - f_{\kappa_\nu} F_{\kappa_x} S_{0JJ}(-\kappa_\nu, -\kappa_x)]  \right. \\
        &\left. - g_A \sqrt{\frac{2J+1}{4\pi}} \delta \rho^{(i,f)}_{\boldsymbol{C}_J \gamma_5}(\boldsymbol{r}) [g_{\kappa_\nu} F_{\kappa_x} S_{0 JJ}(\kappa_\nu, -\kappa_x) + f_{\kappa_\nu} G_{\kappa_x} S_{0JJ}(-\kappa_\nu, \kappa_x)]  \right. \\
         &\left. + g_V \sqrt{\frac{2L+1}{4\pi}} \delta \rho^{(i,f)}_{[\boldsymbol{C}_L \otimes \boldsymbol{\alpha}]_J}(\boldsymbol{r}) [g_{\kappa_\nu} F_{\kappa_x} S_{1LJ}(\kappa_\nu, -\kappa_x) + f_{\kappa_\nu} G_{\kappa_x} S_{1LJ}(-\kappa_\nu, \kappa_x)]  \right. \\  
         &\left. + g_A \sqrt{\frac{2L+1}{4\pi}} \delta \rho^{(i,f)}_{[\boldsymbol{C}_L \otimes \boldsymbol{\Sigma}]_J}(\boldsymbol{r}) [g_{\kappa_\nu} G_{\kappa_x} S_{1LJ}(\kappa_\nu, \kappa_x) - f_{\kappa_\nu} F_{\kappa_x} S_{1LJ}(-\kappa_\nu, -\kappa_x)]  \right\},
    \end{split}
\end{align}
\end{widetext}
where $g_V$ and $g_A$ are the vector and axial-vector coupling constants, respectively. The normalized spherical harmonic is defined as $C_{LM} = \sqrt{\frac{4 \pi}{2L+1}} Y_{LM}$, $\boldsymbol{\alpha} = \gamma_0 \boldsymbol{\gamma}$ is the Dirac matrix, and $\boldsymbol{\Sigma} = \boldsymbol{\alpha} \gamma_5$ is the Pauli spin matrix embedded in the Dirac space, namely $\boldsymbol{\Sigma} = \begin{pmatrix}
    \boldsymbol{\sigma} & 0 \\
    0 & \boldsymbol{\sigma}
\end{pmatrix}$.
Furthermore, we use the abbreviation of  Ref. \cite{Horiuchi2021a} for the  angular wave function:
\begin{equation}
    S_{K L J}(\kappa^\prime, \kappa) = \sqrt{2}\hat{j}_\kappa \hat{j}_{\kappa^\prime} \hat{l}_\kappa \hat{l}_{\kappa^\prime} \hat{K} C^{L 0}_{l_\kappa 0 l_{\kappa^\prime} 0} \begin{Bmatrix}
        l_{\kappa^\prime} & 1/2 & j_{\kappa^\prime} \\
        l_\kappa & 1/2 & j_\kappa \\ 
        L & K & J
    \end{Bmatrix},
\end{equation}
where $\hat{j} = \sqrt{2j+1}$. Since  the covariant structure of the Dirac spinors for nuclear states is preserved, no further non-relativistic reduction of matrix elements is performed, and they can be directly coupled with the lepton spinors in Eq. (\ref{eq:matrix_element_ERWF}).

\begin{figure}
    \centering
    \includegraphics[width=0.8\linewidth]{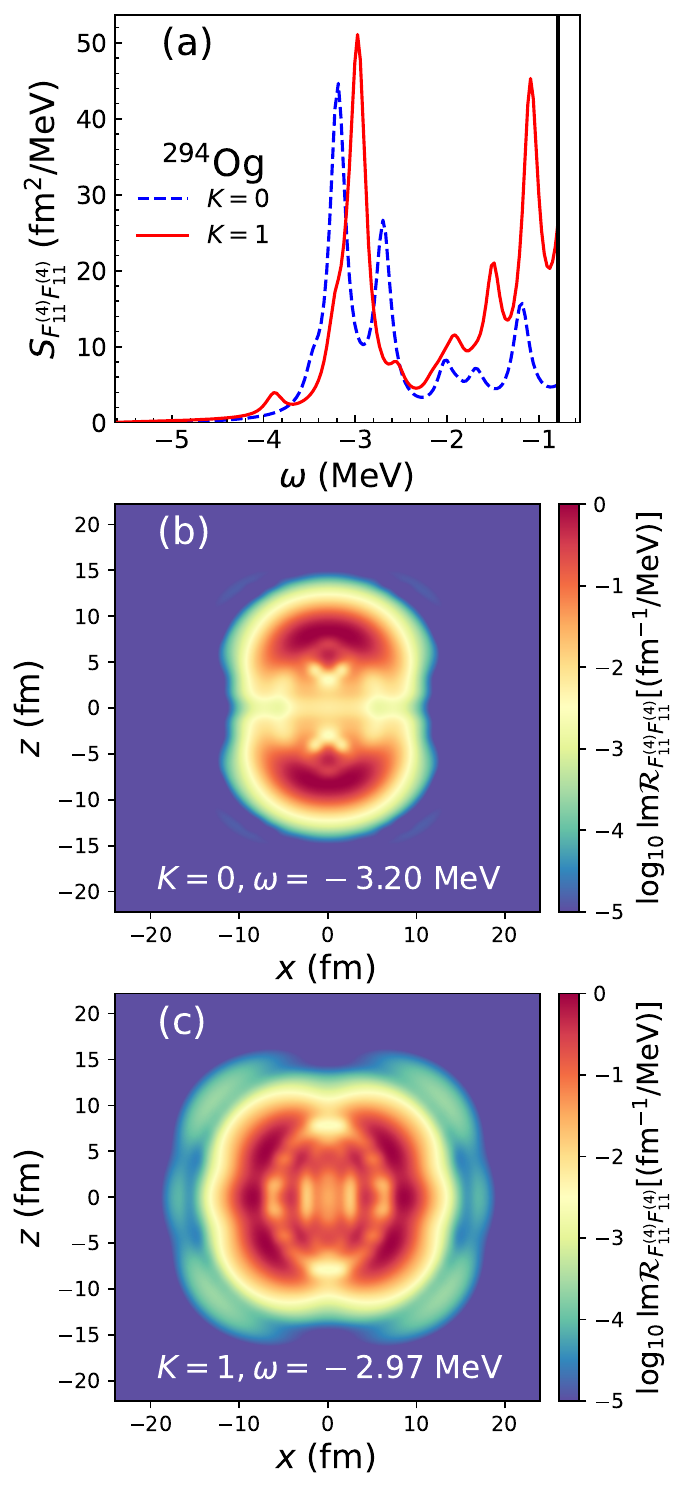}
    \caption{(a) The $1^-$ spin-dipole strength function, $S_{F_{11}^{(4)}F_{11}^{(4)}}$, in ${}^{294}$Og within the $Q_{EC}$ energy window for $K = 0$ (dashed line) and $K = 1$ (solid line) projections. Also shown is the imaginary part of the integrated response $\mathcal{R}_{F_{11}^{(4)}F_{11}^{(4)}}$ for $K = 0$ peak at $\omega = -3.20$ MeV (b) and $K = 1$ peak at $\omega = -2.97$ MeV (c).}
    \label{fig:transition_density}
\end{figure}

\subsection{EC rate within the pnRQRPA}
Next, we extend the formulation of the linear response pnRQRPA to explicitly couple transition densities to the radial parts of lepton wave functions.
Within the linear response, Eq. (\ref{eq:rate_equation}) has the following form for the decay of the ground state of the  even-even parent nucleus:
\begin{align}\label{eq:EC_rate_QRPA}
\begin{split}
   \lambda_x &= \frac{G_F^2 V_{ud}^2}{2\pi} \sum \limits_\mu (E_0(\Omega_\mu) + E_x)^2 \\
   &\times \sum \limits_{\kappa_\nu} \sum \limits_{JL}  \left| \langle \mu || \Xi_{JL}(\kappa_x, \kappa_\nu) || 0 \rangle \right|^2,  
\end{split}
\end{align}
where the summation over initial and final states $(i,f)$ is replaced with summation over pnRQRPA poles with eigenvalues $\Omega_\mu$. The end-point energy $E_0^{(i,f)}$ is replaced with $E_0(\Omega_\mu)$ defined as
\begin{equation}\label{endpoint}
    E_0(\Omega_\mu) = \Omega_\mu - \lambda_p + \lambda_n - \Delta_{nH},
\end{equation}
where $\lambda_{n(p)}$ is the neutron(proton) chemical potential, and $\Delta_{nH} = 0.782$ MeV is the mass difference between the neutron and hydrogen atom. It is to be noted that 
within the linear response pnRQRPA \cite{Ravlic2021a,Ravlic2024a}, one replaces the sum over  discrete poles by the integral of the   strength function.

The sum over excited states in Eq. (\ref{eq:EC_rate_QRPA})
\begin{equation}\label{eq:transition_density_rate}
\sum \limits_\mu (E_0(\Omega_\mu) + E_x)^2 \sum \limits_{J L\kappa_\nu} |\langle \mu || \Xi_{JL}(\kappa_x, \kappa_\nu) || 0 \rangle|^2,
\end{equation}
can be rewritten in terms of transition densities [cf. Eq. (\ref{eq:transition_density})]:
\begin{align}
\begin{split}
 &|\langle \mu || \Xi_{JL}(\kappa_x, \kappa_\nu) || 0 \rangle|^2 = \\
 &\left| \sum \limits_{n =1}^4 \int d\boldsymbol{r} \delta \rho_{F^{(n)}_{JL}}(\Omega_\mu, \boldsymbol{r}) \mathcal{L}_{\kappa_x \kappa_\nu}^{(n)}(\Omega_\mu,\boldsymbol{r}) \right|^2,
 \end{split}
\end{align}
where $\mathcal{L}_{\kappa_x \kappa_\nu}^{(n)}(\Omega_\mu,\boldsymbol{r}) $ labels the lepton part of four terms in Eq. (\ref{eq:matrix_element_ERWF}), while $ \delta \rho_{F^{(n)}_{JL}}(\Omega_\mu, \boldsymbol{r})$ is the transition density containing the nuclear part. The operators 
$F^{(n)}_{JL}$ are explicitly defined as:
\begin{align}
\begin{split}
    &\hat{F}^{(1)}_{JL=0} = \boldsymbol{C}_J, \quad \hat{F}^{(2)}_{JL=0} = \boldsymbol{C}_J \gamma_5, \\
    &\hat{F}^{(3)}_{JL} = [\boldsymbol{C}_L \otimes \boldsymbol{\alpha}]_J, \quad \hat{F}^{(4)}_{JL} = [\boldsymbol{C}_L \otimes \boldsymbol{\Sigma}]_J. \\
\end{split}
\end{align}
The expression for the matrix element $\Xi_{JL}$ can also be evaluated in the complex plane by integrating over suitably chosen contour $\mathcal{C}$, encircling the QRPA pole at $\Omega_\mu$
\begin{equation}\label{eq:response_def}
|\langle \mu | \hat{F}_{JL} | 0 \rangle|^2 = \int d\boldsymbol{r} \int d\boldsymbol{r}^\prime \frac{1}{2 \pi i } \oint_\mathcal{C} d\omega \hat{R}_{F_{JL}F_{JL}}(\omega, \boldsymbol{r},\boldsymbol{r}^\prime),
\end{equation}
where the response function $\hat{R}_{F_{JL}F_{JL}}(\omega, \boldsymbol{r},\boldsymbol{r}^\prime)$ is defined as
\begin{align}
\begin{split}
    \hat{R}_{F_{JL}F_{JL}}(\omega, \boldsymbol{r}, \boldsymbol{r}^\prime) &= \sum \limits_\mu \frac{ \delta \rho_{F_{JL}}(\Omega_\mu, \boldsymbol{r})\delta \rho^*_{F_{JL}}(\Omega_\mu, \boldsymbol{r}^\prime)}{\omega - \Omega_\mu }  \\
    &-  \frac{\delta \rho^*_{F_{JL}}(\Omega_\mu, \boldsymbol{r})\delta \rho_{F_{JL}}(\Omega_\mu, \boldsymbol{r}^\prime)}{\omega + \Omega_\mu }.
\end{split}
\end{align}
That is,  the matrix element can be obtained as the residue of the response function:
\begin{equation}
|\langle \mu | \hat{F}_{JL} | 0 \rangle|^2 \sim
\int \int d\boldsymbol{r} d\boldsymbol{r}^\prime \text{Res}[\hat{R}_{F_{JL}F_{JL}}(\omega, \boldsymbol{r}, \boldsymbol{r}^\prime), \Omega_\mu] .
\end{equation}
The response function satisfies the Bethe-Salpeter  equation:
\begin{align}\label{BS}
\begin{split}
  &\hat{R}_{F_{JL}F_{JL}}(\omega, \boldsymbol{r}, \boldsymbol{r}^\prime) =  \hat{R}^0_{F_{JL}F_{JL}}(\omega, \boldsymbol{r}, \boldsymbol{r}^\prime) \\ 
  &+ \int \int d \boldsymbol{r}^{\prime \prime} d \boldsymbol{r}^{\prime \prime \prime} \hat{R}^0_{F_{JL}F_{JL}}(\omega, \boldsymbol{r}, \boldsymbol{r}^{\prime \prime}) v(\boldsymbol{r}^{\prime \prime}, \boldsymbol{r}^{\prime \prime \prime}) \\
  &\times \hat{R}_{F_{JL}F_{JL}}(\omega, \boldsymbol{r}^{\prime \prime \prime}, \boldsymbol{r}^{\prime}),
  \end{split}
  \end{align}
where $\hat{R}^0_{F_{JL}F_{JL}}$ is the unperturbed response, and $v$ is the residual interaction. Details on solving Eq.~(\ref{BS})  can be found in Refs. \cite{Ravlic2021a,Ravlic2024a}. The definition of the response function can be generalized to include different operators $F^{(i)}_{JL}$ and $F^{(j)}_{JL}$ so that Eq. (\ref{eq:response_def}) becomes
\begin{align}
\begin{split}
&\langle \mu | \hat{F}^{(i)}_{JL} | 0 \rangle \langle \mu | \hat{F}^{(j)}_{JL} | 0 \rangle^*  = \\
&\int d\boldsymbol{r} \int d\boldsymbol{r}^\prime \frac{1}{2 \pi i } \oint_\mathcal{C} d\omega \hat{R}_{F^{(i)}_{JL} F^{(j)}_{JL}}(\omega, \boldsymbol{r},\boldsymbol{r}^\prime),
\end{split}
\end{align}
which allows us to deal with the interference terms. The matrix element of $\Xi_{JL}$ contains a sum of different operators, such as axial and vector components of the weak-interaction Hamiltonian, and leads to interference terms between axial and vector terms. We define the 
singly-integrated response function on a radial mesh as
\begin{equation}
 \mathcal{R}_{F^{(i)}_{JL}F^{(j)}_{JL}}(\omega, \boldsymbol{r}) = \int d \boldsymbol{r}^\prime \hat{R}_{F^{(i)}_{JL} F^{(j)}_{JL}}(\omega, \boldsymbol{r}, \boldsymbol{r}^\prime),  
\end{equation}
and the corresponding doubly-integrated response function as
\begin{equation}\label{eq:doubly-integrated_response}
 R_{F^{(i)}_{JL}F^{(j)}_{JL}}(\omega) = \int d \boldsymbol{r} \int d \boldsymbol{r}^\prime \hat{R}_{F^{(i)}_{JL} F^{(j)}_{JL}}(\omega, \boldsymbol{r}, \boldsymbol{r}^\prime),  
\end{equation}
both of which will be used in the following.

The strength function $S_{F_{JL}F_{JL}}(\omega)$, an experimental observable, is defined as the imaginary part of the doubly-integrated response function
\begin{equation}
    S_{F^{(i)}_{JL}F^{(j)}_{JL}}(\omega) = -\frac{1}{\pi } \text{Im} R_{F^{(i)}_{JL}F^{(j)}_{JL}}(\omega ).
\end{equation}

In Fig.~\ref{fig:transition_density}(a) we show the strength function  for the $1^-$ spin-dipole mode, $S_{F^{(4)}_{11} F^{(4)}_{11}}$, in ${}^{294}$Og defined by the operator
\begin{equation}
    \hat{F}^{(4)}_{J=1,L=1} = [ \boldsymbol{C}_1 \otimes \boldsymbol{\Sigma} ]_1,
\end{equation}
within the $Q_{EC}$ window. A small imaginary value $\eta = 0.1$ MeV is added to energy, $\omega \to \omega + i \eta$, to get the finite Lorentzian width of the peaks. We note that our calculations predict ${}^{294}$Og having a deformed  oblate shape in the ground state. For systems with axial deformation, instead of total angular momentum $J$, a good quantum number is its projection on the $z$-axis, $K$. In 
Figs.~\ref{fig:transition_density}(b) and (c) we show the imaginary part of the integrated response function $\mathcal{R}_{F^{(4)}_{11}F^{(4)}_{11}}(\omega, \boldsymbol{r})$, for particular peaks corresponding to $K = 0$ and $K = 1$, respectively. We note that the integrated response function $\mathcal{R}_{F^{(4)}_{11}F^{(4)}_{11}}$ can help us visualize different excitation modes of the nucleus at a given energy. In particular, we observe that the $K = 0$ mode corresponds to excitations along the $z$-axis, while the $K = 1$ mode represents excitations in the perpendicular direction.

\begin{figure*}
\includegraphics[width=0.92\linewidth]{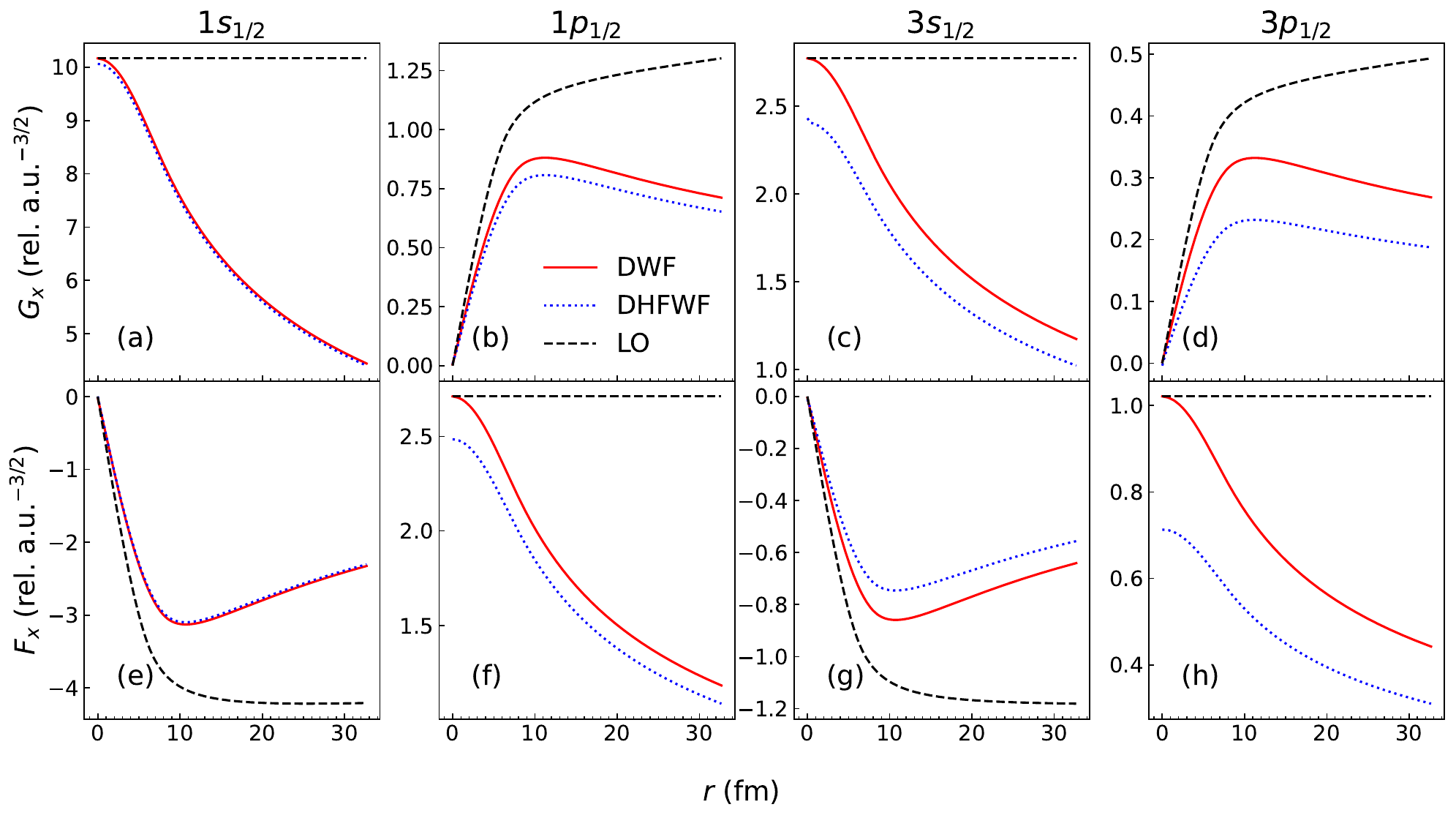}
\caption{Comparison between different calculation methods used to obtain bound-state electron radial wave functions for specific orbitals in ${}^{294}$Og. The LO approximation of Eq.~(\ref{eq:electron_LO}) (dashed line) is compared with single-particle Dirac (DWF) calculations (solid line) and Dirac-Hartree-Fock (DHFWF) calculation (dotted line). Panels (a)--(d) display upper components $G_x$ while (e)--(h) show lower components $F_x$.  The relativistic atomic units are defined such that $\hbar = c = m_e = 1$.}\label{fig:ERWFs}
\end{figure*}

In terms of the response function $\hat{R}_{FF}(\omega, \boldsymbol{r},\boldsymbol{r}^\prime)$, Eq. (\ref{eq:transition_density_rate}) can be rewritten as
\begin{align}\label{eq:full_rate}
\begin{split}
&\sum \limits_\mu (E_0(\Omega_\mu) + E_x)^2 \sum \limits_{J L\kappa_\nu} |\langle \mu || \Xi_{JL}(\kappa_x, \kappa_\nu) || 0 \rangle|^2  \\
&=  \sum \limits_{J L \kappa_\nu}\int d\boldsymbol{r} \int d\boldsymbol{r}^\prime \frac{1}{2\pi i} \oint_{\mathcal{C}} d \omega (E_0(\omega) + E_x)^2 \\
&\times \sum \limits_{i,j = 1}^4 \hat{R}_{F^{(i)}_{JL}F^{(j)}_{JL}}(\omega, \boldsymbol{r}, \boldsymbol{r}^\prime) \mathcal{L}_{\kappa_x \kappa_\nu}^{(i)}(\omega,\boldsymbol{r}) \mathcal{L}_{\kappa_x \kappa_\nu}^{(j)*}(\omega,\boldsymbol{r}^\prime).
\end{split}
\end{align}
This allows us to calculate the required nuclear and lepton matrix elements within the pnRQRPA formulation. However, we note that the integration over contour $\mathcal{C}$ requires analytic continuation of all previously defined functions of excitation energy $\omega$ in the complex plane. In particular, this concerns neutrino and electron wave functions. Analytical continuation is straightforward for spherical Bessel functions, while the bound-state ERWFs should not be impacted by the continuation, since they are independent of the nuclear excitation energy.

\subsection{Reduction to the LO ERWFs}

To compare our results with the existing literature, and study the impact of advanced lepton wave functions on the EC rate, it is convenient to perform the reduction of the transition matrix element in Eq. (\ref{eq:matrix_element_ERWF}) to the LO. In the LO approximation, the ERWFs in Eq. (\ref{eq:electron_wavefunction}) can be expanded, for $\kappa_x = \mp 1$, as defined in Ref. \cite{Bambynek1977a}:
\begin{align}\label{eq:electron_LO}
\begin{split}
& G_{-1}^{LO} = \beta_{-1}, \\
&F_{-1}^{LO}(r) = - \beta_{-1}r \left[ \frac{E_x - m_e}{3} + \xi \frac{2}{3}I(1,1,1,1;r) \right], \\
& G_1^{LO}(r) = \beta_1 r \left[ \frac{E_x+m_e}{3}+\xi \frac{2}{3} I(1,1,1,1;r) \right], \\
&F_1^{LO} = \beta_1,
\end{split}
\end{align} 
where $\beta_{1} = F_1(0)$, and $\beta_{-1} = G_{-1}(0)$ are the Coulomb amplitudes, $\xi = \alpha Z/(2R)$, where $Z$ is charge number, and $R$ is the nuclear root-mean-square radius. In the above expressions, $I(|\kappa_x|,1,1,1;r)$ is the function that takes into account the finite size of the nuclear charge distribution. For the homogeneously charged sphere, this function is \cite{Bambynek1977a}:
\begin{align}\label{eq:I_function}
\begin{split}
&I(|\kappa_x|,1,1,1;r) = \left[
\frac{3}{2} - \frac{2|\kappa_x|+1}{2(2|\kappa_x|+3)} \left( \frac{r}{R} \right)^2 \right] \theta(R-r) \\
&+ \left[ \frac{2|\kappa_x|+1}{2|\kappa_x|} \frac{R}{r} - \frac{3}{2|\kappa_x|(2|\kappa_x|+3)}\left(\frac{R}{r} \right)^{2|\kappa_x|+1} \right]\theta(r-R),
\end{split}
\end{align}
where $\theta(x)$ is the Heaviside step function. The $\kappa_\nu = \mp 1$ neutrino wave functions in the LO become:
\begin{align}\label{eq:neutrino_LO}
\begin{split}
&g_{-1}^{LO} = 1, \quad f_{-1}^{LO}(r) = -\frac{k_\nu r}{3}, \\
&g_1^{LO}(r) = \frac{k_\nu r}{3}, \quad f_1^{LO}  = 1.
\end{split}
\end{align}
Before reducing Eq. (\ref{eq:rate_equation}) to the LO limit, it is instructive to compare the LO ERWFs with the corresponding single-particle Dirac ERWFs (DWF), calculated within the RADIAL solver, and the DHF ERWFs (DHFWF) calculated with GRASP. 
In Fig. \ref{fig:ERWFs} we compare the ERWFs in the lowest lying orbitals $1s_{1/2}$ and $1p_{1/2}$ of ${}^{294}$Og, as well as somewhat higher lying $3s_{1/2}$ and $3p_{1/2}$ orbitals. The  LO is a  crude approximation to ERWFs. Furthermore, we observe that the difference between single-particle DWFs and DHFWFs is negligible for the low-lying $1s_{1/2}$ state, as it is well screened from other electrons by the large nuclear charge. However, the differences between DWF and DHFWF results increase significantly as the electronic binding energy decreases, and electron-electron interactions become more important.

In the following, we perform the LO limit of the first-forbidden (FF) $1^-$ multipole, which  has the largest contribution to the total EC rate in SHN as demonstrated in Ref. \cite{Ravlic2024b}. At the LO, the partial waves that contribute the most are $(\kappa_x, \kappa_\nu) = (\pm 1, \mp 1)$. For the space-like vector component, we retain only $L = 0$ (cf. third term in Eq. (\ref{eq:matrix_element_ERWF})), while for the space-like axial component, we keep $L = 1$ [cf. fourth term in Eq. (\ref{eq:matrix_element_ERWF})]. Based on this $J=1, L$ assignment, the EC matrix element reduces to:
\begin{align}
         &\langle i || \Xi_{1 1}( 1, - 1) || f \rangle \notag \\ 
         &\approx  \int d\boldsymbol{r} \left\{ - g_V \sqrt{\frac{3}{4 \pi}} \delta \rho^{(i,f)}_{\boldsymbol{C}_1}(\boldsymbol{r}) \frac{\sqrt{6}}{3} [g_{- 1} G_{ 1} + f_{- 1} F_{ 1}] \right. \notag \\
         &+ g_V \frac{1}{\sqrt{4 \pi}} \delta \rho^{(i,f)}_{[\boldsymbol{\alpha}]_1}(\boldsymbol{r})[\sqrt{2} g_{- 1} F_{ 1}]   \\
         & \left. + g_A \sqrt{\frac{3}{4\pi}} \delta \rho^{(i,f)}_{[\boldsymbol{C}_1 \otimes \boldsymbol{\Sigma}]_1}(\boldsymbol{r}) \frac{2\sqrt{3}}{3} [g_{- 1} G_{+ 1} - f_{- 1} F_{+ 1}] \right\}, \notag
\end{align}
and
\begin{align}
         &\langle i || \Xi_{1 1}(- 1, 1) || f \rangle \notag \\ 
         &\approx  \int d\boldsymbol{r} \left\{ + g_V \sqrt{\frac{3}{4 \pi}} \delta \rho^{(i,f)}_{\boldsymbol{C}_1}(\boldsymbol{r}) \frac{\sqrt{6}}{3} [g_{ 1} G_{- 1} + f_{ 1} F_{- 1}] \right. \notag \\
         &+ g_V \frac{1}{\sqrt{4 \pi}} \delta \rho^{(i,f)}_{[\boldsymbol{\alpha}]_1}(\boldsymbol{r})[\sqrt{2} f_{ 1} G_{- 1}]   \\
         & \left. + g_A \sqrt{\frac{3}{4\pi}} \delta \rho^{(i,f)}_{[\boldsymbol{C}_1 \otimes \boldsymbol{\Sigma}]_1}(\boldsymbol{r}) \frac{2\sqrt{3}}{3} [g_{ 1} G_{- 1} - f_{ 1} F_{- 1}] \right\}, \notag
\end{align}
for $\kappa_x = \pm 1$, and $\kappa_\nu = \mp 1$, respectively.
In the above, we neglect all $\mathcal{O}(r^2)$ terms representing the multiplication of two wave functions that already contain a power of $r$. Using the notation of Ref. \cite{Marketin2016a} for transition densities: 
\begin{align}\label{eq:operators_LO}
    \begin{split}
        & g_V \int d \boldsymbol{r} \boldsymbol{r} \delta \rho^{(i,f)}_{\boldsymbol{C}_1}(\boldsymbol{r}) = -x, \\
        & g_V \int d \boldsymbol{r} \boldsymbol{r} I(1,1,1,1;r) \delta \rho^{(i,f)}_{\boldsymbol{C}_1}(\boldsymbol{r}) = - \frac{3}{2}x^\prime, \\
        & g_V \int d \boldsymbol{r} \delta \rho^{(i,f)}_{[\boldsymbol{\alpha}]_1}(\boldsymbol{r}) = - \xi^\prime y, \\
        & g_A \int d \boldsymbol{r} \boldsymbol{r} \delta \rho^{(i,f)}_{[\boldsymbol{C}_1 \otimes \boldsymbol{\Sigma}]_1}(\boldsymbol{r}) = -\frac{1}{\sqrt{2}} u, \\
        & g_A \int d \boldsymbol{r} \boldsymbol{r} I(1,1,1,1;r) \delta \rho^{(i,f)}_{[\boldsymbol{C}_1 \otimes \boldsymbol{\Sigma}]_1}(\boldsymbol{r}) = -\frac{3}{2\sqrt{2}} u^\prime, \\
    \end{split}
\end{align}
the matrix element becomes:
\begin{align}\label{eq:simple_matrix_element}
    \begin{split}
         &\langle i || \Xi_{1 1}(\pm 1,\mp 1) || f \rangle =  - \frac{1}{\sqrt{2 \pi}} \beta_{\pm 1} \xi^\prime y \\ 
         &+ \frac{1}{\sqrt{2 \pi}} \beta_{\pm 1} \left[ \frac{1}{3} (E_x \pm m_e - k_\nu)x + \xi x^\prime \right] \\
         &- \frac{1}{\sqrt{2 \pi}} \beta_{\pm 1} \left[ \frac{1}{3}(E_x \pm m_e + k_\nu)u + \xi u^\prime \right].
    \end{split}
\end{align}
This expression agrees with the LO  formulas presented in  Refs. \cite{Bambynek1977a,Behrens1970a,Behrens1971a, Marketin2016a}.  

The matrix elements induced by the operators in Eq. (\ref{eq:operators_LO}) can be represented in terms of doubly-integrated response function $R_{FG}$ [cf. Eq. (\ref{eq:doubly-integrated_response})], where in general $F \neq G$.
 To exemplify the calculation of $1^-$ rate in the LO, we can introduce the {shape-factor} $C_x$, which has the following form:
 \begin{equation}\label{eq:shape-factor}
     C_x(\omega) = ka_x(\omega) + 2 \text{sgn}(\kappa_x)kb_x(\omega) + kc_x(\omega), \quad \kappa_x = \pm 1,
 \end{equation}
obtained by appropriately grouping the terms in Eq. (\ref{eq:simple_matrix_element}), with expressions for $ka_x, kb_x$, and $kc_x$ detailed in Appendix \ref{sec:appa}.

 Finally, the expression for the EC rate in the LO approximation becomes:
\begin{equation}\label{eq:LO_rate}
    \lambda_{EC} = \frac{1}{2\pi i} \frac{\text{ln}2}{K} \sum \limits_x n_x \oint_{\mathcal{C}} d \omega C_x(\omega) f_x(W_0[\omega]),
\end{equation}
where $f_x = \frac{\pi }{2} \beta_x^2 (W_0[\omega] + W_x)^2$ is the phase-space factor, $W_x = E_x/(m_ec^2)$, $n_x$ is the relative occupation of electrons in a given shell, and the end-point energy is defined as
\begin{equation}
    W_0[\omega](m_e c^2) = \omega -\lambda_p + \lambda_n - \Delta_{nH},
\end{equation}
cf. Eq.~(\ref{endpoint}). In the above, $K$ is the super-allowed $\beta$-decay constant \cite{Marketin2016a}.

\begin{figure}
    \centering
    \includegraphics[width=0.9\linewidth]{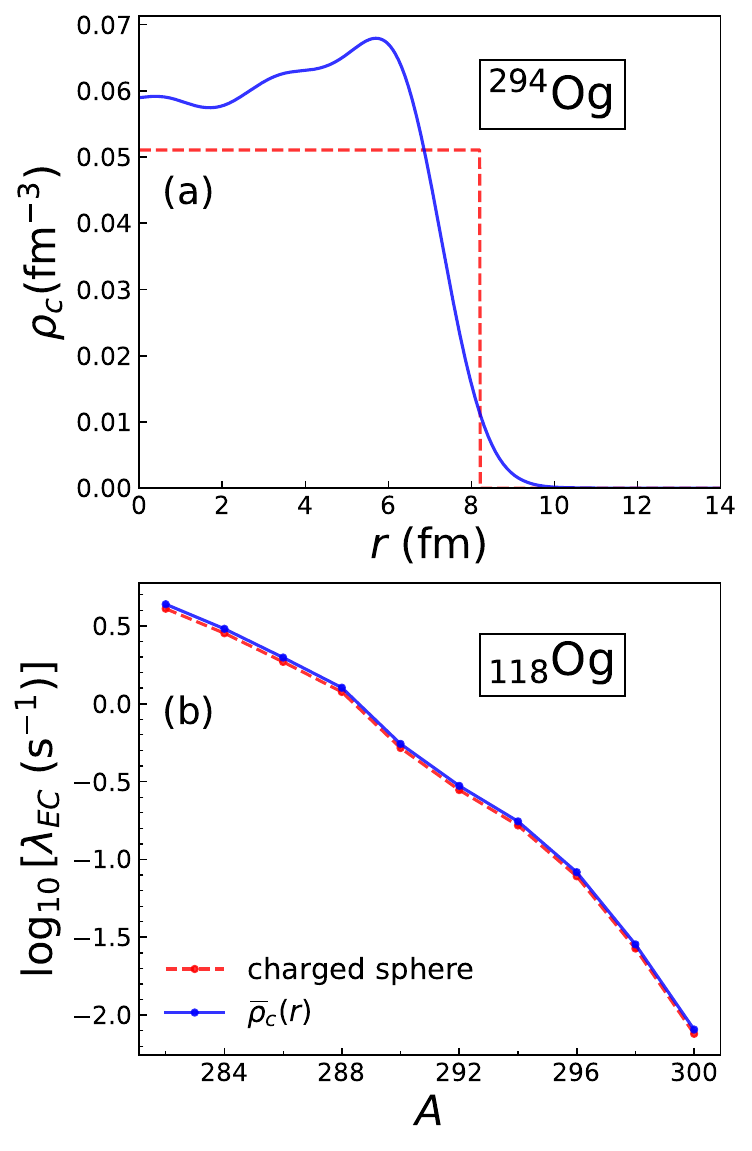}
    \caption{(a) The nuclear charge density $\rho_c$ in ${}^{294}$Og computed by assuming a homogeneously charged sphere with radius $R$ (dashed line) or directly from nuclear DFT calculation (solid line). (b) The isotopic dependence of EC rates for $1^-$ FF transition in oganesson computed using the two prescriptions for charge density.}
    \label{fig:charge_density}
\end{figure}

 \begin{figure*}[ht!]
    \includegraphics[width=0.9\linewidth]{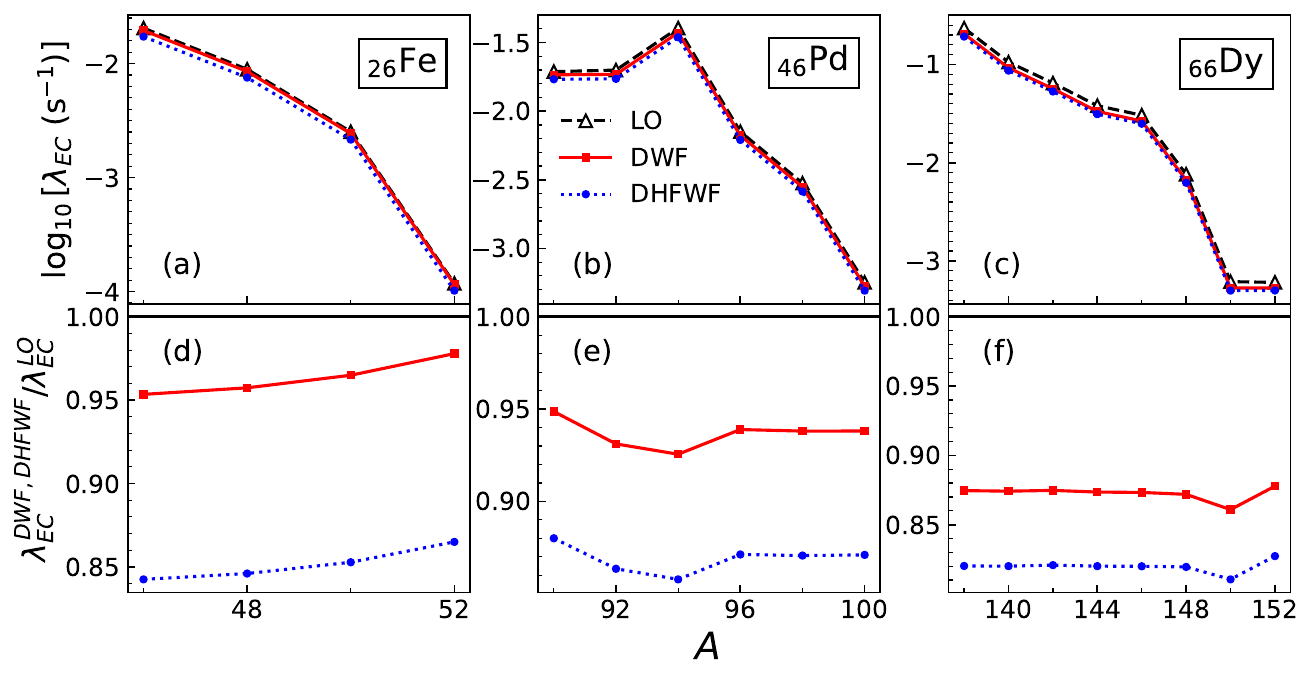}
    \caption{Top: The electron capture rate $\lambda_{EC}$ for isotopic chains of (a) iron, (b) palladium, and (c) dysprosium calculated using DHFWF (circles), DWF (squares), and LO (triangles) ERWFs. Bottom: The ratio between EC rates calculated with either DHFWF or DWF and ERWFs obtained within the LO approximation.}
    \label{fig:LO_vs_exact_GT_only}
\end{figure*}

\section{Results}\label{sec:results}
In our  linear response pnRQRPA calculations, we use $N_{osc} = 20$ oscillator shells in a stretched axially-deformed basis and $N_{GH} = N_{GL} = 25$ points for coordinate-space integration. The calculation employs the DD-PC1 EDF, with the charge-exchange residual interaction couplings calibrated in Ref. \cite{Ravlic2024b}.  The complex integration in Eqs. (\ref{eq:full_rate}) and (\ref{eq:LO_rate}) is performed by discretizing a circular contour using a 60-point Gauss-Legendre quadrature. While  the present study is restricted  to even-even parent nuclei,  our conclusions apply to odd-$A$ and odd-odd parent nuclei as well. In the light and medium-mass nuclei, in which protons and neutrons occupy similar orbitals, and chemical potentials are fairly similar, EC is dominated by the allowed  GT transitions. In heavy nuclei, however, in which protons and neutrons occupy different shells, and  hence there is a large asymmetry between chemical potentials, the forbidden transitions dominate as the allowed Gamow-Teller transitions are Pauli-blocked \cite{Langanke2003,Ney2020a,Kumar2024}. In particular, the EC rate in SHN is dominated by the FF $J^\pi = 1^-$ transitions, with a minor contribution of GT \cite{Ravlic2024b}. Therefore, both allowed and forbidden transitions are considered in our calculation. The electron/neutrino relativistic block numbers are constrained to $\kappa_{x(\nu)} = \pm 1, \pm 2$, which provides excellent convergence in the  mass regions studied. The major radial number for electron states is $n \leq 4$. For each of the 4 terms in the expression for the transition rate (\ref{eq:rate_equation}) we retain the lowest non-vanishing $L$. The second term, corresponding to the time-like axial component is neglected, as is the exchange-overlap correction between parent and daughter states as its effect is relatively small \cite{Bambynek1977a}. In the calculations of ERWFs, the nuclear charge distribution is taken as a homogeneous sphere with charge $Z$ and radius corresponding to the nuclear root-mean-square radius $R$ calculated within DFT. We note that EC rates are expected to be fairly independent of the details of nuclear charge distribution up to $Z \sim 170$ \cite{Smits2023a,Smits2024a}. To explicitly check for the impact of using realistic nuclear charge distributions, we compare the EC rate calculations based on two treatments of nuclear charge densities. In the first approach, we approximate the charge density as a homogeneous sphere with the radius $R = \sqrt{\frac{5}{3}} \sqrt{\langle r^2 \rangle}$, where the root-mean-square radius $\sqrt{\langle r^2 \rangle}$ is calculated within the nuclear DFT. In the second case, we employ the realistic charge densities calculated with nuclear DFT to generate the Coulomb potential. Starting from the charge density $\rho_c(\boldsymbol{r})$, the Coulomb potential can be obtained by direct integration
\begin{equation}
    V_c(\boldsymbol{r}) = - \int d \boldsymbol{r}^\prime \frac{\rho_c(\boldsymbol{r}^\prime)}{|\boldsymbol{r} - \boldsymbol{r}^\prime|}.
\end{equation}
Since we assume axially-deformed density distribution, a simplification can be made by integrating over the angle $\theta$ to obtain the averaged density
\begin{equation}
    \bar{\rho}_c(r) = \frac{Z}{2 \mathcal{N}}\int \limits_0^\pi \rho_c(r\sin \theta, r \cos \theta) \sin \theta d \theta,
\end{equation}
where $\mathcal{N} = 4\pi \int \limits_0^\infty r^2 \bar{\rho}_c(r) dr$ is the norm introduced to preserve the particle number.

In both cases, to calculate the EC rate we employ DWFs by making necessary adjustments to the RADIAL solver \cite{Salvat1995a}. In Fig. \ref{fig:charge_density}(a) we compare the charge densities $\bar{\rho}_c$ calculated with both methods. It is interesting to notice a central depression in DFT charge density, a consequence of the Coulomb frustration \cite{Schuetrumpf2017a}. The homogeneous sphere charge density with effective radius $R$ represents the leading monopole correction to the full charge density. Finally, in Fig. \ref{fig:charge_density}(b) we show the FF $1^-$ EC rate in the isotopic chain of oganesson ($Z = 118$).
The uniformly-charged sphere approximation to the full charge density suffices to accurately model the EC decay in SHN. The ratio of EC rates between the two methods is at most 6\%.

Before further extending our analysis to SHN, we benchmark our calculations in the regions of nuclei with lower
atomic numbers where it is expected that detailed distribution of lepton wave functions will not have a large effect on the EC rates.
Therefore, we first  investigate three isotopic chains: ${}_{26}$Fe, ${}_{46}$Pd and ${}_{66}$Dy.  The GRASP calculations are performed by assigning the relativistic electronic ground-state configurations according to  Tab. \ref{tab:FeDyPd}. Since, compared to the nuclear transition matrix elements, the ERWFs do not vary much across the isotopic chain, we use one representative nucleus for which we perform RADIAL and GRASP calculations. These are: ${}^{50}$Fe, ${}^{96}$Pd, and ${}^{146}$Dy.

\begin{table}[htb]
    \caption{Relativistic atomic configurations in the ground-state used for the GRASP calculations with the corresponding coupling to total angular momentum $J$.}\label{tab:FeDyPd}
    \begin{ruledtabular}
    \begin{tabular}{c|c|c}
       atom & configuration & $J$ \\  
    \hline
        $_{26}$Fe & [Ar]$3d_{3/2}^4 3d_{5/2}^2 4s_{1/2}^2$ & 4  \\
        $_{46}$Pd & [Kr]$4d_{3/2}^4 4d_{5/2}^6$ & 0  \\
        $_{66}$Dy & [Xe]$4f_{7/2}^8 4f_{5/2}^2 6s_{1/2}^2$ & 4  \\
    \end{tabular}
    \label{tab:my_label}
    \end{ruledtabular}
\end{table}

\begin{figure}[htb]
    \includegraphics[width=\linewidth]{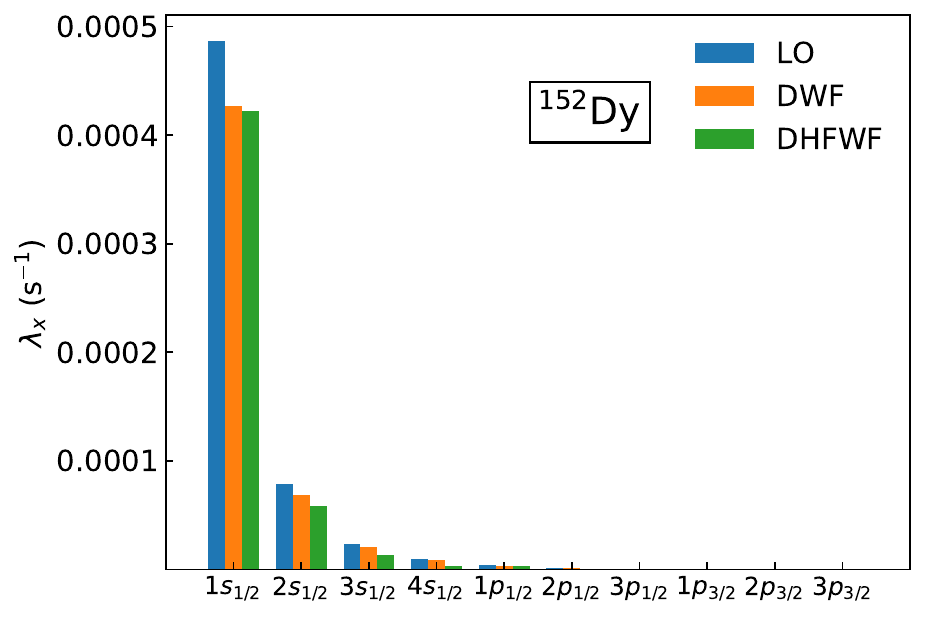}
    \caption{The partial contribution $\lambda_x$ of orbital $x$ to the total EC rate in ${}^{152}$Dy. Comparison is made between DWF, DHFWF and LO calculations.}
    \label{fig:Dy152_x}
\end{figure}

The isotopic dependence of EC rates is shown in Fig. \ref{fig:LO_vs_exact_GT_only}(a)-(c).  The EC rates in these nuclei are dominated by allowed GT transitions, while the FF ones are suppressed by orders of magnitude, and therefore neglected. It is seen that
going beyond the LO has an overall small effect that tends to increase with $Z$. In general,  improving the  ERWFs results in a rate reduction, see Fig. \ref{fig:LO_vs_exact_GT_only}(d)-(f). For the DWFs, the reduction is around 5\% in iron, and increases to around 12\% in dysprosium, varying only slightly throughout the isotopic chain. For the DHFWFs,  the reduction of EC rates in iron tends to be around 15\%, and increases to almost 20\% in dysprosium. Furthermore, one notices that as the nuclear charge increases, the difference between EC rates calculated using DWFs and DHFWFs decreases. 
Such an effect is a result of the screening of inner orbitals by the nuclear charge. Figure~\ref{fig:Dy152_x} shows the contribution of individual orbitals $x$ to the total EC rate in ${}^{152}$Dy. By far the largest  contribution comes from the orbitals with $\kappa_x = -1$, and in particular the innermost $1s_{1/2}$. However, due to its relative proximity to the nucleus, the $1s_{1/2}$ orbital is partially screened from the other electrons in the atom by the nuclear charge.

 \begin{figure}[htb]
     \includegraphics[width=0.8\linewidth]{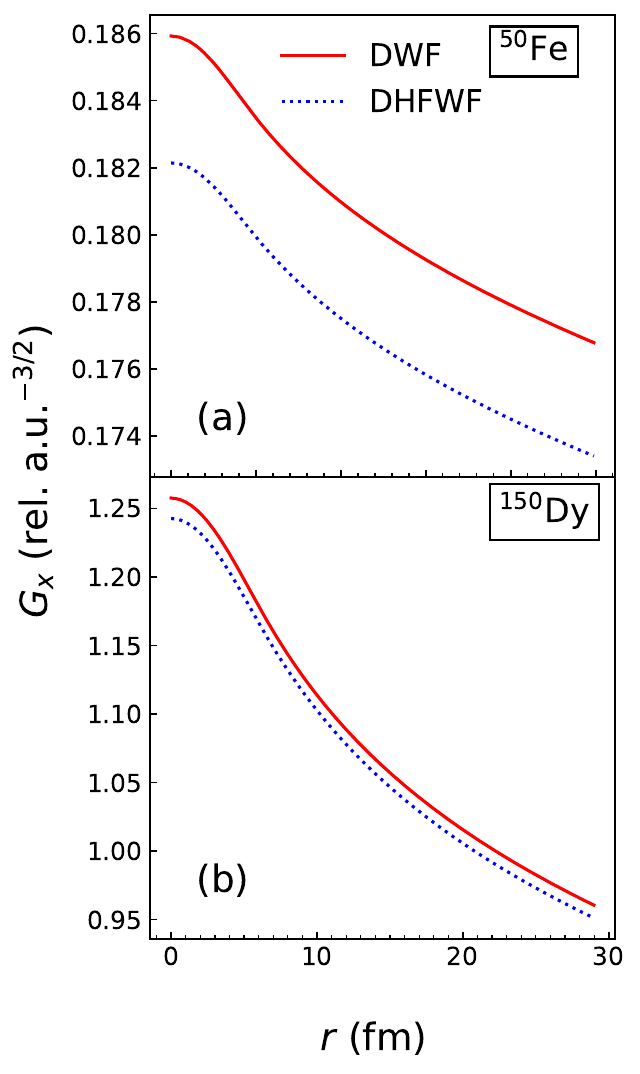}
     \caption{The comparison between the upper components $G_x$ of the $1s_{1/2}$ wave functions calculated within single-particle Dirac (DWF) (solid line) and Dirac-Hartree-Fock (DHFWF) method (dotted line) in ${}^{50}$Fe (top) and ${}^{150}$Dy (bottom).}
     \label{fig:wf_Fe_Dy}
 \end{figure}

 The influence of the screening can be best seen in Fig. \ref{fig:wf_Fe_Dy} that shows the upper component $G_x$ of the $1s_{1/2}$ spinor, for ${}^{50}$Fe (a), and ${}^{150}$Dy (b). For ${}^{50}$Fe, DHF correlations reduce the wave function by 2\%, around the origin. However, for ${}^{150}$Dy, with $Z = 66$, this reduction is around 1\%. Certainly a small change, but with some impact on EC rates.
We note that the proton-rich nuclei in the considered isotopic chains are dominated by $\beta^+$-decay rather than EC. Such a calculation would require handling the electron wave functions in the continuum and is left for future work.

\begin{figure}[htb]
    \includegraphics[width=\linewidth]{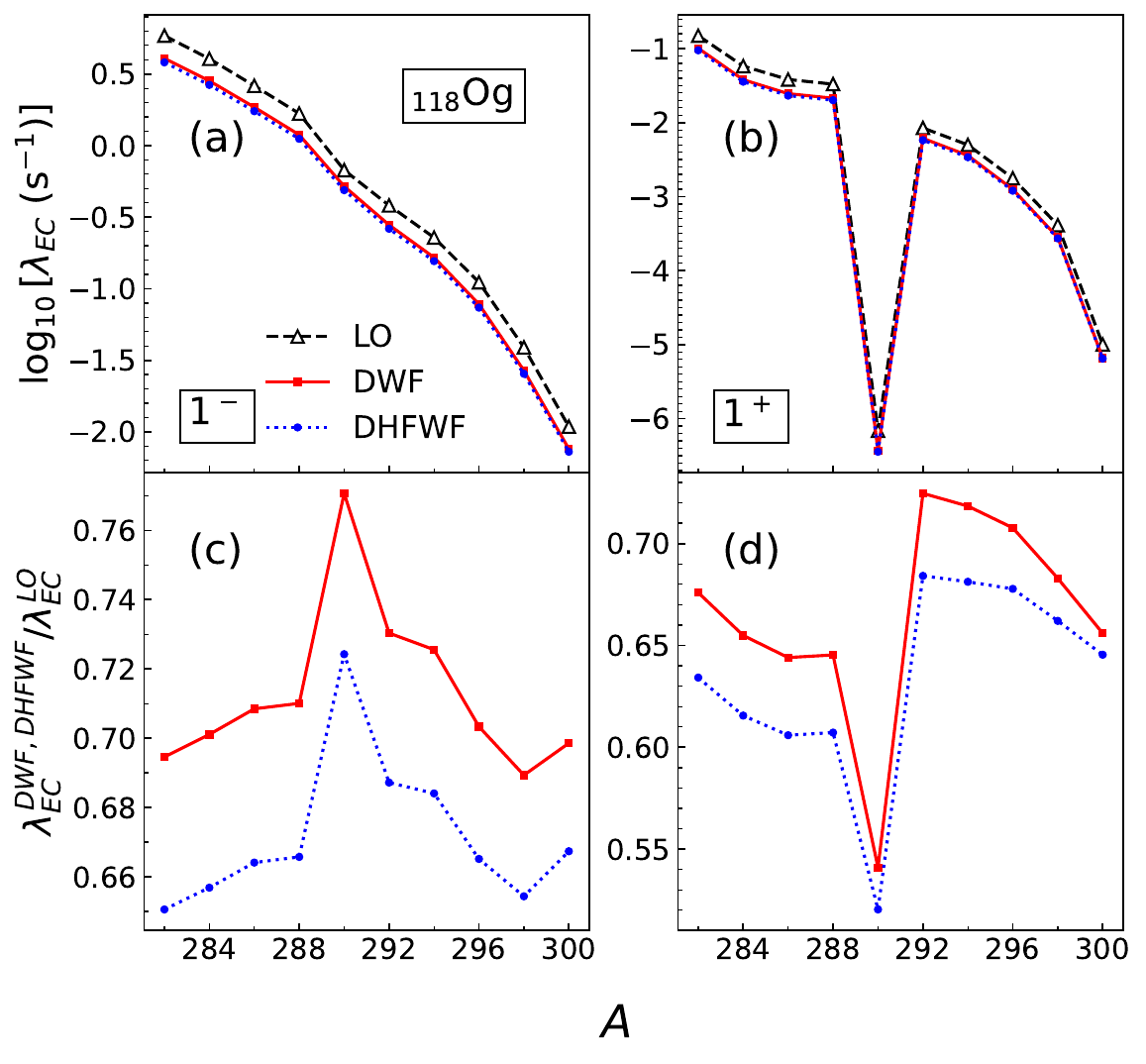}
    \caption{Top: The electron capture rate $\lambda_{EC}$ for even-even nuclei in the isotopic chain ${}_{118}$Og for (a) FF $1^-$ and (b) GT $1^+$ multipoles. The rate is calculated with DWF (squares), DHFWF (circles) and the LO approximation (triangles). Bottom: The ratio between EC rates calculated with DWF, DHFWF, and LO, for (c) $1^-$  and (d) $1^+$ multipoles.}
    \label{fig:Og_rate_EC}
\end{figure}

\begin{figure}[htb]
    \includegraphics[width=\linewidth]{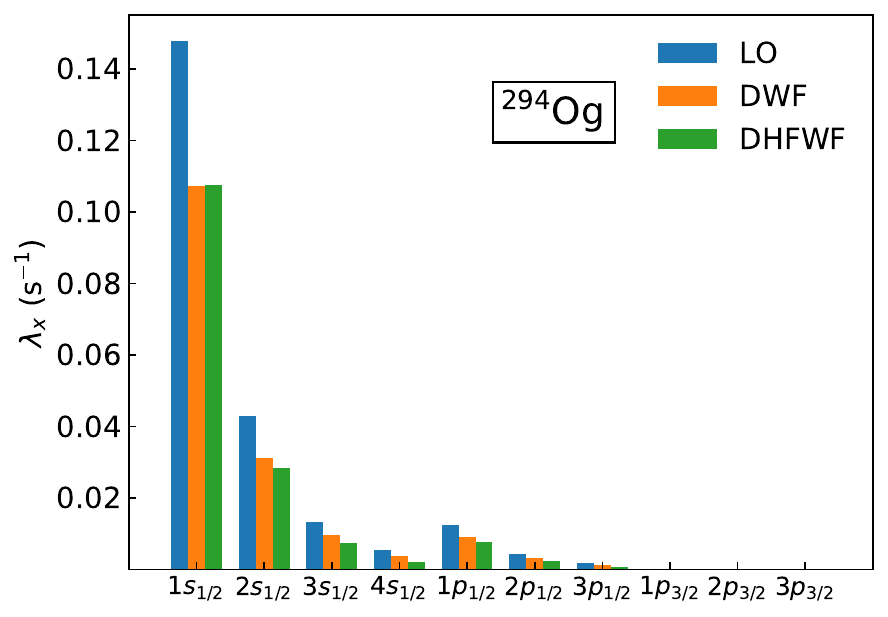}
    \caption{Similar as in Fig. \ref{fig:Dy152_x} but for ${}^{294}$Og.}
    \label{fig:Og294_orbital_contribution}
\end{figure}

Next, we extend our analysis to SHN. By considering only one CSF, obtaining a ground-state atomic configuration of superheavy atoms would require minimizing the ground-state energy over a set of possible CSFs by solving the variational DHF equations. However, by considering closed electronic shell configurations, we avoid calculations of additional configurations, as there is only one CSF to consider, and the calculation reduces to the DHF solved by the variational method. Therefore, in the following analysis, we choose the isotopic chain of ${}_{118}$Og. In Ref. \cite{Ravlic2024a}, we have demonstrated that EC dominates over $\beta^+$-decay in SHN. Therefore, the calculations presented here completely determine the weak-decay rate of SHN. We perform calculations for the FF $1^-$, which dominates the rate, and the allowed GT $1^+$ transitions. The EC rates for the even-even oganesson isotopes are shown in Fig. \ref{fig:Og_rate_EC}. Compared to lighter systems,   the differences between LO and more sophisticated ERWFs are more pronounced, for both $1^-$ and $1^+$ multipoles, with the main effect  being the reduction of the rates. The  $1^-$ transitions are predicted to have a larger rate by almost two orders of magnitude compared to the GT transitions. Although the $1^-$ rates tend to decrease smoothly with increasing neutron number, the GT rates display an interesting kink for $^{290}$Og. Our calculations, based on the DD-PC1 EDF, predict that the ground state of ${}^{290}$Og is spherical. Nuclei with lower mass numbers in the isotopic chain have a prolate shape, while nuclei with higher mass transition to oblate shapes. Due to the great disparity between proton and neutron Fermi levels for spherical ${}^{290}$Og, the GT rate is significantly hindered due to the Pauli blocking. Indeed, deformation effects in SHN help to surmount the large gap between Fermi levels. By taking the ratio between the EC rates with either DWFs or DHFWFs, and the LO approximation, a similar pattern is obtained as discussed previously in Fig. \ref{fig:LO_vs_exact_GT_only}. In particular, for $1^-$, we observe that the DWFs tend to reduce the rate by around 30\%, while for DHFWFs the reduction is around 34\%. The $1^+$ multipole displays a similar reduction. As expected, the effect of going beyond the LO approximation is more pronounced in SHN. However, it is interesting to notice that taking into account electron-electron interactions within the DHFWFs, further reduces the rate by only 4\%. For comparison, in the iron isotopic chain, this number was around 10\%.  Again, this is mainly due to the screening of the innermost electrons by the large nuclear charge. To illustrate this point, in Fig. \ref{fig:Og294_orbital_contribution}, we show the partial contribution of electron orbitals to the total EC rate in ${}^{294}$Og. The main contribution again stems from the $\kappa_x = -1$ orbitals, and in particular $1s_{1/2}$. One can also notice an increased contribution from the $\kappa_x = +1$ orbitals, especially $1p_{1/2}$, with all others being negligible. By inspecting the $1s_{1/2}$ wave function in Fig. \ref{fig:ERWFs}(a), it can be seen that DWFs and DHFWFs strongly overlap. For the upper component $G_x$, the ratio of the wave function around the origin is sub-1\% for both calculations, clearly demonstrating a strong screening effect.

\section{Conclusions}

In this work, we investigated the sensitivity of EC rates in medium-mass and superheavy nuclei on the fidelity of lepton wave functions. To this end, we carried out  quantified predictions of EC rates using  state-of-the-art nuclear and atomic models. The nuclear response functions were calculated using the covariant nuclear DFT within the pnRQRPA assuming axially-deformed nuclei. Two approaches were employed to obtain the electron wave functions: (i) a single-particle electron picture, where wave functions were obtained numerically with the RADIAL code, and (ii)  GRASP DHF calculations. Both approaches are compared to the standard LO approximation  often used in weak-decay estimates. For the medium-mass  isotopic chains, the allowed transitions dominate the rate as protons and neutrons move in similar shells. In this case, compared to the LO approximation, the single-particle DWFs tend to reduce the rates from 5\% in Fe isotopic chain, up to 12\% in Dy isotopes. The DHFWFs yield even larger reductions in rates: up to 15\% is Fe chain and up to 20\% in Dy. 
For nuclei where precise EC data are available, details of lepton wave functions could be crucial in successfully reproducing and predicting  experimental half-lives.

For the superheavy nuclei, represented by the Og chain, the forbidden transitions dominate due to the Pauli blocking. Here, more sophisticated electron wave functions can lead to a reduction in EC rate of up to 40\%. Interestingly, compared to DWF, including electron-electron interactions within the DHFWF leads to only a minor reduction. 
This is explained in terms of screening effects of the nuclear charge on the innermost electron orbitals, in particular $1s_{1/2}$.

\section{Acknowledgements}
This work was supported by the U.S. Department of Energy under Award Number DOE-DE-NA0004074 (NNSA, the Stewardship Science Academic Alliances program) and by the Office of Science, Office of Nuclear Physics under grants DE-SC0013365 and DE-SC0023175 (Office of Advanced Scientific Computing Research and Office of Nuclear Physics, Scientific Discovery through Advanced Computing). This work was also supported in part through computational resources and services provided by the Institute for Cyber-Enabled Research at Michigan State University.

\appendix
\section{Detailed expression for shape-factor in LO}\label{sec:appa}
Based on the LO expansion of neutrino and electron wave functions, the shape-factor $C_x$ can be written in the form defined in Eq. (\ref{eq:shape-factor}), where:

\begin{align}
 \begin{split}
         &ka_x(\omega) = R_{(\xi^\prime y)(\xi^\prime y)}(\omega) + \xi^2 R_{(x^\prime)(x^\prime)}(\omega) \\ 
         &+ 1/9W_0^2R_{(x)(x)}(\omega) + \xi^2 R_{(u^\prime)(u^\prime)}(\omega) \\
         &+ 1/9(W_x + k_\nu)^2 R_{(u)(u)}(\omega) + 2\xi R_{(x^\prime)(\xi^\prime y)}(\omega) \\
        &-2/3 W_0 R_{(x)(\xi^\prime y)}(\omega) -2 \xi R_{(u^\prime)(\xi^\prime y)}(\omega)  \\
        &-2/3 (W_x + k_\nu) R_{(u)(\xi^\prime y)}(\omega) -2/3 W_0 \xi R_{(x)(x^\prime)}(\omega) \\
        &- 2 \xi^2 R_{(u^\prime)(x^\prime)}(\omega) -2/3 \xi (W_x+k_\nu) R_{(u)(x^\prime)}(\omega) \\
        &+ 2/3 \xi W_0 R_{(u^\prime)(x)}(\omega) + 2/9 W_0 (W_x+k_\nu) R_{(u)(x)}(\omega) \\
        &+ 2/3 \xi (W_x+k_\nu) R_{(u)(u^\prime)}(\omega), \\
\end{split}
\end{align}
\begin{align}
    \begin{split}
        &kb_x(\omega) = 1/9  R_{(x)(x)}(\omega) - 2/9 R_{(u)(x)}(\omega)  \\
         &+ 1/9 R_{(u)(u)}(\omega), \\
    \end{split}
\end{align}

\begin{align}
    \begin{split}
        &kc_x(\omega) = +1/3 R_{(x)(\xi^\prime y)}(\omega) -1/3 R_{(u)(\xi^\prime y)}(\omega)  \\
         &+ 1/3 \xi R_{(x)(x^\prime)}(\omega) -1/3 \xi R_{(u)(x^\prime)}(\omega) \\
         &- 1/9 W_0 R_{(x)(x)}(\omega) + 1/9 W_0 R_{(u)(x)}(\omega) \\
        &-1/3 \xi R_{(u^\prime)(x)}(\omega) + 1/3 \xi R_{(u)(u^\prime)}(\omega) \\
        &- 1/9 (W_x + k_\nu) R_{(u)(x)}(\omega) +1/9 (W_x + k_\nu) R_{(u)(u)}(\omega).
    \end{split}
\end{align}

 For instance, $R_{(u)(u^\prime)}(\omega)$ is the doubly-integrated response function [cf. Eq. (\ref{eq:doubly-integrated_response})] induced by two external field operators $\hat{F} = \sqrt{2}g_A r [ \boldsymbol{C}_1 \otimes \boldsymbol{\sigma} ]_1$, and $\hat{G} = \frac{2\sqrt{2}}{3}g_A I(1,1,1,1;r)r [ \boldsymbol{C}_1 \otimes \boldsymbol{\sigma} ]_1$.
 In total, for $1^-$, 15 response functions have to be calculated, including the interference terms.
 \clearpage

%

\end{document}